\documentclass[11pt,a4paper]{article}
\usepackage{jcappub}
\title{Gravitational Wave from Axion-SU(2) Gauge Fields: Effective Field Theory for Kinetically Driven Inflation}
\author[a,b]{Yuki Watanabe}
\author[a,c]{and Eiichiro Komatsu}
\affiliation[a]{Max Planck Institute for Astrophysics, Karl-Schwarzschild-Str.1, 85741 Garching, Germany}
\affiliation[b]{Department of Physics, National Institute of Technology, Gunma College, Gunma 371-8530, Japan}
\affiliation[c]{Kavli Institute for the Physics and Mathematics of the Universe, WPI, UTIAS, The University of Tokyo, Chiba 277-8583, Japan}
\emailAdd{yuki.watanabe@gunma-ct.ac.jp}
\emailAdd{komatsu@mpa-garching.mpg.de}
\abstract{%
Building on Weinberg's approach to effective field theory for inflation, we construct an effective Lagrangian for a pseudo scalar (axion) inflaton field with shift symmetry. In this Lagrangian we allow the axion field to couple to non-Abelian gauge fields via a Chern-Simons term. We then analyze a class of inflation models driven by kinetic terms. We find that the observational constraints on the amplitudes of curvature perturbations and non-Gaussianity yield a lower bound for the tensor-to-scalar ratio of $r\gtrsim 5\times 10^{-3}$ from the vacuum fluctuation. The sourced gravitational wave from SU(2) gauge fields further increases the tensor-to-scalar ratio and makes the total gravitational wave partially chiral and non-Gaussian, which can be probed by polarization of the cosmic microwave background and direct detection experiments. We discuss constraints on parameter space due to backreaction of spin-2 particles produced by the gauge field. 
}
\begin{document}
\maketitle
\section{Introduction}
Phenomenological success of cosmic inflation \cite{Sato:1980yn,Guth:1980zm} requires a flat potential for a slowly-rolling scalar field $\phi$ \cite{Albrecht:1982wi,Linde:1981mu}. Since the seminal work by Freese, Frieman and Olinto \cite{Freese:1990rb}, shift symmetry, symmetry under a constant shift of $\phi\to \phi+c$, has often been used to construct the necessary flat potential. In this setup, a pseudo Nambu-Goldstone boson, an axion field, is identified as the inflaton field, and the flat potential emerges as a consequence of softly broken shift symmetry (e.g., by the instanton effect).

Another approach is to drive inflation by kinetic terms, ${\cal L}=K(\phi)X+L(\phi)X^2+\dots$, with $X\equiv -(\partial\phi)^2/2$ \cite{ArmendarizPicon:1999rj}. While $X$ is shift symmetric, the coefficients $K$ and $L$ may not be. Nevertheless, we can demand softly broken shift symmetry by requiring $K$ and $L$ to depend on $\phi$ only weakly.

In this paper, we construct an effective Lagrangian for a pseudo scalar field with shift symmetry. The basic idea follows from Weinberg's effective field theory for inflation \cite{Weinberg:2008hq}; namely, the number of spacetime derivatives is less than or equal to four. We then retain terms that are shift symmetric. A novel feature of our Lagrangian is that we also add the shift symmetric Chern-Simons coupling to (non-Abelian) gauge fields, $\phi{F\tilde{F}}$ \cite{Anber:2009ua,Adshead:2012kp}. We find that our construction predicts a {\it lower} bound for the tensor-to-scalar ratio of $r\gtrsim 5\times 10^{-3}$ with partially chiral and non-Gaussian gravitational waves. 
In our setup, shift symmetry breaking effects can explain the tilt of the scalar curvature power spectrum, $n_{\rm s}<1$, discovered by the cosmic microwave background (CMB) experiments \cite{Hinshaw:2012aka,Komatsu:2014ioa,Akrami:2018odb}.

The rest of this paper is organized as follows. In section~\ref{sec:eft}, we explain our construction of the effective Lagrangian for a pseudo scalar field with shift symmetry. In section~\ref{sec:bg}, we derive the background equations of motion for the scalar and gauge fields and find approximate solutions with softly broken shift symmetry.
In section~\ref{sec:pert}, we analyze the scalar and tensor perturbations and calculate observables such as the scalar spectral tilt and non-Gaussianity as well as the tensor-to-scalar ratio, chiraity, and non-Gaussianity of the primordial gravitational wave. In section~\ref{sec:back}, we constrain the model parameter space using sizes of backreaction of spin-2 particles produced by the gauge field.
We conclude in section~\ref{sec:conclusion}.

\section{Effective field theory for inflation with shift symmetry}\label{sec:eft}
We start with the kinetic Lagrangian with no more than four spacetime derivatives
\cite{ArmendarizPicon:1999rj,Weinberg:2008hq}:
\begin{align}\label{lag0}
{\cal L}_0 = \sqrt{-g}\left[
 (M_{\rm p}^2/2)R + a_1 X + a_2 X^2  \right] \ ,
\end{align}
where $g \equiv \det{g_{\mu\nu}}$, $X \equiv -\partial_{\mu}\phi\partial^{\mu}\phi/2$, $R$ is the Ricci scalar, $M_{\rm p}=(8\pi G)^{-1/2}$ is the reduced Planck mass, and $a_1$ and $a_2$ are coefficients characterized by the mass scale $M$ of theory. 
To achieve inflation, we need $a_1<0$ and $a_2 >0$. Then the cosmic expansion rate $H\equiv \dot{a}/a$ ($a$ is the cosmic scale factor and an over-dot is a time derivative) is given by $M_{\rm p}^2H^2 \sim a_2X^2 \sim M^4$ and $\epsilon \equiv -\dot{H}/H^2 = 0$; a phase of the exact de Sitter inflation is realized without a potential \cite{ArmendarizPicon:1999rj, Garriga:1999vw,ArkaniHamed:2003uy, ArkaniHamed:2003uz}. The configuration $a_1 <0$ and $a_2>0$ has been used to achieve ``ghost inflation" \cite{ArkaniHamed:2003uz} by forming ``ghost condensate" \cite{ArkaniHamed:2003uy}.
If $a_1 <0$ and $a_2<0$, the system is unstable since the Hamiltonian is not positive-definite.

Any additional derivatives acting on $\partial_{\mu}\phi$ or on the metric yield factors of order $H \sim M^2/M_{\rm p} \ll M$, which guarantees that (\ref{lag0}) is the leading terms in the low-energy effective field theory for inflation, and any correction terms are suppressed by factors of $H/M$.
The leading correction to (\ref{lag0}) consists of a sum of all generally covariant  and shift symmetric terms with four spacetime derivatives. It can be put in the form
\begin{align}\label{lag1}
\Delta {\cal L}_1 = \sqrt{-g} \left[ 
a_3 X\Box \phi 
 + a_4 G^{\mu\nu}\partial_{\mu}\phi\partial_{\nu}\phi  
 + a_5 \phi R_{\rm GB} 
\right] 
+ a_6  \phi \epsilon^{\mu\nu\rho\sigma}R_{\mu\nu}{}^{\kappa\lambda}R_{\rho\sigma\kappa\lambda} \ ,
\end{align}
where $\Box \phi \equiv g^{\mu\nu}\nabla_{\mu}\nabla_{\nu}\phi$, $G^{\mu\nu}\equiv R^{\mu\nu}-\frac12 g^{\mu\nu}R$ is the Einstein tensor, $R_{\rm GB} \equiv R^2 -4R_{\mu\nu}R^{\mu\nu}+R_{\mu\nu\rho\sigma}R^{\mu\nu\rho\sigma}$ is the Gauss-Bonnet scalar, and $\epsilon^{\mu\nu\rho\sigma}$ is the totally antisymmetric tensor density with $\epsilon^{0123}\equiv +1$.
The coefficients $a_i$ ($i = 3, 4, 5, 6$) are characterized by the mass scale $M$ (or higher scale like $M_{\rm p}$ depending on underlying ultra-violet theory).

The correction terms given in (\ref{lag1}) are the most general ones with four spacetime derivatives \cite{Weinberg:2008hq}. The other terms such as $(\Box\phi)^2$, $R^{\mu\nu}\partial_{\mu}\phi\partial_{\nu}\phi$, $R\Box\phi$, and so on, are eliminated by the field equations. 
We can estimate sizes of the correction terms as $a_3X\Box\phi \sim HM^3$, $a_4G^{\mu\nu}\partial_{\mu}\phi\partial_{\nu}\phi\sim H^2M^2$, $a_5\phi R_{\rm GB} \sim H^3M$, and $a_6 \phi \epsilon^{\mu\nu\rho\sigma}R_{\mu\nu}{}^{\kappa\lambda}R_{\rho\sigma\kappa\lambda}\sim H^3M$, with $X\sim M^4$ and $\phi \sim \dot\phi\Delta t \sim M^2/H$.
Since terms with six spacetime derivatives like $(\Box\phi)^3$ would be on the order of $H^3M$, we  ignore the last two terms in (\ref{lag1}); moreover, they can be rewritten as bilinear in the Weyl tensor that vanishes at the background level in a conformally flat spacetime \cite{Weinberg:2008hq}. 

The first three terms in (\ref{lag1}) introduce no auxiliary field (known as the Ostrogradski ghost mode) and  theory is stable on a cosmological background \cite{Horndeski:1974wa,Deffayet:2011gz,Kobayashi:2011nu}, while the last term in (\ref{lag1}) causes instability if ${\cal L}_0 + \Delta{\cal L}_1$ is taken as the full Lagrangian \cite{Crisostomi:2017ugk}.
The last term is parity violating in a nontrivial background of $\phi$; thus, it results in chiral gravitational waves with different amplitudes of right- and left-handed helicities \cite{Lue:1998mq}. Chirality can be as large as several tens of percent when the cut-off scale of theory, $M$, is as low as $M=20H$ \cite{Mirzagholi:2020irt}; the effect becomes smaller for a larger cut-off (e.g., Planck scale \cite{Alexander:2004wk,Kamada:2019ewe}). 
Specifically, we assume $M\sim \sqrt{HM_{\rm p}}$ in this paper. The (perturbative) strong coupling scales in the gravity-scalar sector were studied in \cite{Kunimitsu:2015faa} where they found $M=\epsilon^{1/4}\sqrt{HM_{\rm p}}$ if (\ref{lag0}) dominates among others.

Finally, we add another shift symmetric term to the Lagrangian, a Chern-Simons coupling term between $\phi$ and gauge fields. In this paper we focus on SU(2) gauge fields, as they (or a SU(2) subgroup of non-Abelian gauge fields) acquire an isotropic and homogeneous background solution during inflation when conformal invariance is broken by a four derivative operator $(F\tilde{F})^2$ \cite{Maleknejad:2011jw, Maleknejad:2011sq} or by a Chern-Simons interaction $\phi F\tilde{F}$ \cite{Adshead:2012kp}, where $F$ is the gauge field strength tensor. The former can be obtained as an effective Lagrangian of the latter by integrating out the massive $\phi$ on energy scales below its mass scale \cite{Adshead:2012qe,SheikhJabbari:2012qf,Maleknejad:2012fw}.
 We assume that a global symmetry breaking scale $f$ lies in a range of $H < f < M_{\rm p}$.
The gauge sector is given by
\begin{align}
\Delta{\cal L}_2 = - \frac14 \sqrt{-g}F_{\mu\nu}^aF^{a\mu\nu}
  -\frac{\lambda}{8f}\phi\epsilon^{\mu\nu\rho\sigma}F_{\mu\nu}^aF_{\rho\sigma}^a \ ,
\end{align}
where $F^a_{\mu\nu}\equiv \partial_{\mu}A^a_{\nu}-\partial_{\nu}A^a_{\mu}+g_A\epsilon^{abc}A^b_{\mu}A^c_{\nu}$ is the field strength tensor, $g_A$ is the gauge coupling constant, superscripts $a$, $b$, $c$ are the SU(2) group indices, and summation is assumed for repeated indices. $\lambda$ is a dimensionless coefficient associated to microphysics of the axion \cite{Agrawal:2018mkd}.

 As perturbations in the SU(2) field around the homogeneous and isotropic vacuum expectation value contain tensor modes \cite{Maleknejad:2011jw, Maleknejad:2011sq}, it can source gravitational waves at linear order. The resulting signal is chiral \cite{Adshead:2013qp,Adshead:2013nka,Maleknejad:2012fw,Dimastrogiovanni:2012ew} and non-Gaussian due to self-coupling of SU(2) gauge fields \cite{Agrawal:2017awz,Agrawal:2018mrg,Fujita:2018vmv,Dimastrogiovanni:2018xnn}.\footnote{A similar phenomenology is obtained with a Chern-Simons coupling with a U(1) gauge field \cite{Anber:2009ua}. The sourced gravitational wave is chiral \cite{Sorbo:2011rz} and non-Gaussian because it is sourced {\it non}-linearly by the quadratic term in the stress-energy tensor \cite{Barnaby:2010vf,Barnaby:2011vw,Anber:2012du}.}

In summary, our effective action is given by
\begin{align}\label{action}
S = \int d^4x\sqrt{-g}\left[
 \frac{M_{\rm p}^2}{2}R + a_1 X + a_2 X^2 + a_3 X\Box \phi 
 + a_4 G^{\mu\nu}\partial_{\mu}\phi\partial_{\nu}\phi \right. \nonumber \\  
 \left. - \frac14 F_{\mu\nu}^aF^{a\mu\nu}
  -\frac{\lambda}{8f\sqrt{-g}}\phi\epsilon^{\mu\nu\rho\sigma}F_{\mu\nu}^aF_{\rho\sigma}^a
\right] \ ,
\end{align}
where we have set $a_5 = a_6 =0$ based on the estimates of their magnitudes given above.
If we impose symmetry under parity, the term with $a_3$ is absent.
In the present model, parity symmetry will be spontaneously broken by the Chern-Simons interaction once the gauge field acquires a nontrivial background value. Thus, we consider the case $a_3 \neq 0$ as well.

Here we provide comparison with the previous work on the scalar field Lagrangian non-minimally coupled to gravity. A canonical scalar field corresponds to $a_1=1$. The terms with $a_1$ and $a_4$ are included in ``UV-protected inflation'' \cite{Germani:2010hd,Germani:2011ua} with $a_1=1$ and $a_4=1/(2M^2)$ in their notation. The terms with $a_1$, $a_2$, and $a_3$ are included in ``G-inflation'' \cite{Kobayashi:2010cm} if we set their free functions to $K=a_1X+a_2X^2$ and $G=-a_3X$, which is equivalent to ``kinetic gravity braiding" \cite{Deffayet:2010qz} if $K=a_1X+a_2X^2$ and $G=a_3X$ (or $a_1=1$ and $a_3=-M(\phi)$ in the notation of \cite{Maity:2014oza}).
The terms with $a_1$ and $a_3$ are included in ``galileon inflation" \cite{Burrage:2010cu} if we set their coefficients to $c_2=a_1$ and $c_3 \Lambda^{-3}=-a_3/2$.
In ``generalized G-inflation'' \cite{Kobayashi:2011nu}, our scalar field Lagrangian is realized when $K=a_1X+a_2X^2$, $G_3=-a_3X$, $G_4=M_{\rm p}^2/2$, and $G_5=-a_4\phi$, which is equivalent to $G_4=M_{\rm p}^2/2+a_4X$ up to total derivative. The terms with $a_4$ and $a_5$ are included in ``the Fab Four" \cite{Charmousis:2011bf} if we set their free functions to $V_{\rm george}=M_{\rm p}^2/2$, $V_{\rm john}=a_4$, $V_{\rm ringo}=a_5\phi=0$, and $V_{\rm paul}=0$. 
Our aim in this paper is {\it not} to work with the most general Lagrangian for a scalar field with shift symmetry, but to work with the Lagrangian that is valid in the low-energy effective field theory.

Shift symmetry results in the exact de Sitter expansion, which fails to explain a small but non-zero tilt of the scalar curvature power spectrum \cite{Hinshaw:2012aka,Komatsu:2014ioa,Akrami:2018odb}.
In our setup, shift symmetry may be broken in three ways. The first possibility is to introduce a potential, e.g., 
$V(\phi) = V_{,\phi}\phi$, where $V_{,\phi}$ is nearly constant.
The specific potential form is not important for realizing inflation in our model, since inflation is assumed to be driven by the kinetic terms given in (\ref{lag0}). The operators $\phi$ and $\phi^2$ are protected by nonrenormalization theorem; thus, shift symmetry is softly broken \cite{Burrage:2010cu}. Any periodic potential arisen from the instanton mechanism breaks shift symmetry without receiving large quantum corrections. The second possibility is to introduce weak $\phi$ dependence on the coefficients, $a_i$. 
Since (\ref{lag0}) is the leading part, we shall assume $a_1 = a_1(\phi)$, $a_2 = a_2(\phi)$, $a_3 =$ const., and $a_4 = $ const. for simplicity; thus,
${\cal L}_0 \to \sqrt{-g}\left[(M_{\rm p}^2/2)R +a_1(\phi)X + a_2(\phi)X^2 - V(\phi)\right]$. The third possibility is backreaction of particle production by the gauge field on the equation of motion for $\phi$, which turns out to be too small to break shift symmetry effectively (section~\ref{sec:back}).
In any case, the symmetry breaking terms should be understood as small. 

\section{Inflationary background}\label{sec:bg}
We take the flat, homogeneous and isotropic background such that $ds^2 = -N^2(t)dt^2+a^2(t)d{\bf x}^2$, $\phi = \phi(t)$, $A_0^a = 0$, and $ A_i^a = \delta_i^a a(t)Q(t)$ \cite{Maleknejad:2011jw, Maleknejad:2011sq}.
This homogeneous and isotropic configuration of the gauge field is an attractor solution during inflation \cite{Maleknejad:2013npa,Wolfson:2020fqz}. 
After finding the Hamiltonian constraint (i.e., the Friedmann equation), we set the background lapse function to $N(t) = 1$.

The field equations for $\phi$ and $Q$ are given by
\begin{align}
\dot{J} &+3HJ -K_{,\phi}= -3\frac{g_A\lambda }{f}Q^2(\dot{Q}+HQ) \ ,  \label{eq_phi}\\
J &\equiv \dot\phi (a_1 +a_2\dot{\phi}^2 -3a_3H\dot\phi +6a_4 H^2) \ , \quad
K \equiv  a_1X + a_2X^2 -V\ , \nonumber\\ 
\ddot{Q} &+3H\dot{Q} + (\dot{H}+2H^2)Q +2g_A^2Q^3 = \frac{g_A\lambda}{f}\dot{\phi}Q^2 \ ,\label{eq_Q}
\end{align}
where $K_{, \phi} \equiv \partial K/\partial\phi \ $. The last term in the left hand side of (\ref{eq_Q}) gives an effective mass term for the gauge field background. We write this as $2g_A^2Q^3=2m_Q^2H^2Q$ with $m_Q\equiv g_AQ/H$. 

The flat Friedmann equations are given by
\begin{align}
3M_{\rm p}^2 H^2 &= \rho_{\phi} + \rho_A \ ,\label{rho} \\
-3M_{\rm p}^2H^2 -2M_{\rm p}^2 \dot{H} &= p_{\phi} + p_A \ ,\label{pressure} \\
\rho_{\phi} &= \dot\phi J - K + 6a_4 H^2 X \ , \nonumber\\
p_{\phi} & = K + 2a_3X\ddot{\phi} -6a_4H^2X -4a_4\dot{H}X - 4a_4H\dot{X} \ , \nonumber\\
\rho_A &= \frac32 (\dot{Q} + HQ)^2 + \frac32 g_A^2Q^4 \ , \quad
p_A = \frac13 \rho_A \ .\nonumber
\end{align}
Combining equations~(\ref{rho}) and (\ref{pressure}), we get
\begin{align}\label{epsilon}
\epsilon \equiv -\frac{\dot{H}}{H^2} = \frac{\dot{\phi}J}{2H^2M_{\rm p}^2} +\frac{a_3X\ddot{\phi}}{H^2M_{\rm p}^2}
+2\frac{a_4 X \epsilon}{M_{\rm p}^2} -2\frac{a_4 \dot{X}}{HM_{\rm p}^2} +\frac{2 \rho_A}{3H^2M_{\rm p}^2}\ .
\end{align}
Noting a relation $\epsilon = 3(1+w)/2$ with $w \equiv p/\rho$ being the equation of state ($p$ and $\rho$ are total pressure and energy density, respectively), we would expect $w \simeq 1/3$ and the universe becomes radiation dominated if the last term in (\ref{epsilon}) becomes dominant. However, we will show  in the following that this is not the case for non-vanishing $\lambda$, $g_A $, and $Q$, and the universe is inflationary with $w \simeq -1$ in a quasi-stationary state regardless of the fraction $\rho_A/\rho$. 

We solve the $\phi$ field equation~(\ref{eq_phi}) iteratively using shift symmetry.
Ignoring the $\phi$ dependence (i.e., $K_{,\phi}$), we get the zeroth iterative solution:
\begin{align}\label{sol_j0}
J^{(0)} = -\frac{g_A\lambda}{f}Q^3 + \frac{C}{a^3} \ ,
\end{align}
where the second term is a decaying solution and $C$ is an integration constant.
Here, $J$ is a conjugate momentum of $\phi$ in the absence of the gauge field background $Q$. In the presence of a nontrivial $Q$, the conserved charge associated to shift symmetry is $C = a^3\left(J+\frac{g_A\lambda}{f}Q^3\right)$.
If $Q$ is constant, the solution implies $\dot\phi$ and $H$ are constant and the exact de Sitter expansion ($w = -1$) is realized.

Plugging (\ref{sol_j0}) into (\ref{eq_phi}), we get the first iterative solution:
\begin{align}\label{jsol}
J^{(1)} = -\frac{g_A\lambda}{f}Q^3 +\frac{K_{,\phi} }{3H}+ \frac{C}{a^3} \ ,
\end{align}
where we have assumed that $H$ and $K_{,\phi}$ are nearly constant. The first iterative solution is enough to obtain a quasi de Sitter expansion for a slightly tilted spectrum of curvature perturbations.

A nontrivial, stationary value of $Q$ can be obtained from the $Q$ field equation (\ref{eq_Q}).
To see the appearance of a nontrivial $Q$, we define an effective potential for $Q$:
\begin{align}\label{eq:effpot}
U_{\rm eff}(Q) \equiv \frac12(\dot{H}+2H^2)Q^2 + \frac12 g_A^2Q^4 - \frac{g_A\lambda}{3f}\dot{\phi}Q^3 \ ,
\end{align}
which acquires a nontrivial minimum for $\dot\phi \neq 0$:
\begin{align}\label{qstar}
Q_* =  \frac{\lambda\dot{\phi} }{4g_Af} +  \frac{\lambda}{4g_Af} \sqrt{ \dot\phi^2 -16\frac{f^2H^2(1 -\epsilon/2)}{\lambda^2}} \ , 
\end{align}
where $|\lambda\dot\phi/f| > 4 H$ must be satisfied and $\lambda\dot\phi/(4g_Af) < Q_* < \lambda\dot\phi/(2g_Af)$ for positive values. Equivalently, we can solve the stationary condition $U_{\rm eff}'(Q) = 0$ for $\dot\phi$:
\begin{align}\label{dphistar}
\dot\phi_* = \frac{2fg_AQ}{\lambda}+\frac{2f H^2(1-\epsilon /2)  }{\lambda g_A Q}  \ .
\end{align}

Plugging (\ref{jsol}) and (\ref{dphistar}) into (\ref{epsilon}), we find that the first and last terms in (\ref{epsilon}) nearly cancel, finding a value of $\epsilon$ at stationary trajectory as
\begin{align}\label{epsilonstar}
\left(1-\frac{2a_4X}{M_{\rm p}^2}-\frac{Q^2}{2M_{\rm p}^2}\right)\epsilon_*
= \frac{\dot\phi K_{,\phi}}{6H^3M_{\rm p}^2} +\frac{a_3X\ddot\phi}{H^2M_{\rm p}^2}-\frac{2a_4\dot{X}}{HM_{\rm p}^2} +\frac{\dot{Q}^2}{H^2M_{\rm p}^2}+\frac{2Q\dot{Q}}{HM_{\rm p}^2} \ .
\end{align}
As stated before, quasi de Sitter expansion ($w \simeq -1$) is realized for the attractor solution $\dot\phi = \dot\phi_*$.
In the limit $K_{,\phi}\to 0$, the solutions indicate $\dot\phi \to $ const., $Q \to $ const., and then $\epsilon \to 0$ (i.e., $w \to -1$).
Since we are interested in the quasi-static state, we assume that the field values change slowly as
\begin{align}\label{slow}
\left| \frac{\ddot\phi}{H\dot\phi} \right| \ll 1 \ , \quad
\left| \frac{\dot{Q}}{HQ} \right| \ll 1 \ .
\end{align}  

So far, we have not taken into account backreaction of particle production by the gauge field on the background equations of motion, which modifies the solutions (\ref{jsol}) and (\ref{epsilonstar}). We shall take backreaction into account in section~\ref{sec:back}.

\section{Cosmological perturbations}\label{sec:pert}
The clock of the system is set by the uniform energy density of $\phi$ when $\rho_{\phi} \gg \rho_A$.
In the unitary gauge $\delta\phi = 0$ at all orders, we write the spatial metric as $g_{ij}=a^2e^{2\zeta}[e^{h}]_{ij}=a^2e^{2\zeta}(\delta_{ij}+h_{ij}+h_{ik}h_{kj}/2+\dots)$, where $h_{ij}(t,{\bf x})$ is a symmetric, traceless and divergence-free tensor.

When the effective mass of the gauge field, $m_Q=g_AQ/H$ (\ref{eq_Q}), is small and gauge scalar perturbations do not decouple from the system, instability appears in the scalar perturbation for $m_Q< \sqrt2$ \cite{Dimastrogiovanni:2012ew,Adshead:2013qp}.
Thus, we shall assume $m_Q > \sqrt{2}$ so that gauge scalar perturbations decouple. 

However, tensor perturbations do not decouple and one of the $\pm 2$ helicity states undergoes an exponential amplification due to instability. This tensor mode instability is essential to obtain chiral \cite{Adshead:2013qp,Adshead:2013nka,Maleknejad:2012fw,Dimastrogiovanni:2012ew} and non-Gaussian \cite{Agrawal:2017awz,Agrawal:2018mrg,Fujita:2018vmv,Dimastrogiovanni:2018xnn} gravitational waves.

\subsection{Scalar perturbation: Tilt and non-Gaussianity}
Expanding the action~(\ref{action}) to second order and using the Gauss, Hamiltonian and momentum constraints, we obtain the quadratic action for $\zeta$ at leading order as
\begin{align}
S_{\zeta^2} &= \int d^3xdt \ a^3 {\cal G}_{\rm s} \left[ \dot\zeta^2 - \frac{c_{\rm s}^2}{a^2}(\partial_i \zeta)^2\right] \ ,\\
{\cal G}_{\rm s} &= \frac{\dot\phi}{2H^2}\left[J -6a_3HX +4a_2X\dot\phi\left(1-\frac{2a_3X\dot\phi}{HM_{\rm p}^2}+\frac{8a_4X}{M_{\rm p}^2} \right) \right] \ , \nonumber \\
c_{\rm s}^2 &= \frac{M_{\rm p}^2\epsilon_{\rm s}}{{\cal G}_{\rm s}} \ , \quad
\epsilon_{\rm s} = \epsilon - \frac{a_3X\dot\phi}{HM_{\rm p}^2} +\frac{\xi^2Q^2}{M_{\rm p}^2}\ , \quad
\xi \equiv \frac{\lambda\dot\phi}{2fH} \ ,
\nonumber
\end{align}
where we have used $|a_3X\dot\phi/(HM_{\rm p}^2)|\ll 1$ and $|a_4X/M_{\rm p}^2| \ll 1$ in evaluating ${\cal G}_{\rm s}$ and (\ref{slow}) in $\epsilon_{\rm s}$. Note that ${\cal G}_{\rm s} > 0$ and $c_{\rm s}^2 > 0$ must be satisfied to avoid ghost and gradient instabilities, which requires $a_2 >0$ if the third term in ${\cal G}_{\rm s}$ dominates.
We have estimated the leading contribution from the Gauss constraint (i.e., the equation of motion for non-dynamical field $A_0^a = \delta A_0^a$) as in \cite{Adshead:2013nka}
\begin{align}\label{gauss}
{\cal L}_{(\delta A_0^a)^2} \approx - a \xi^2Q^2(\partial_i\zeta)^2 \ ,
\end{align}
which has yielded the last term in $\epsilon_{\rm s}$ for long-wave modes with $ k \lesssim m_QaH$.\footnote{For short-wave modes, this term contributes as a mass term and suppresses the amplitude of curvature perturbations inside the horizon for $\lambda \gg 1$ \cite{Adshead:2013nka}, which results in the enhancement of the tensor-to-scalar ratio, $r$. In the present case, $\lambda < 1$ and the contribution is negligible. Thus, $r$ is not enhanced for vacuum fluctuations.}
When $\rho_{\phi}\gg \rho_A$, contributions from the gauge field are sub-leading to the Hamiltonian and momentum constraints. Note that the leading part of scalar perturbations is approximated well by a model within refs. \cite{Kobayashi:2010cm, Kobayashi:2011nu} if the gauge field contribution (\ref{gauss}) is ignored. 

For a canonical axion field ($a_1=1$, $a_2=a_3=a_4=0$) with the Chern-Simons coupling,  we obtain ${\cal G}_{\rm s}= M_{\rm p}^2\epsilon$ and $c_{\rm s}^2 \approx 1 +\xi^2 Q^2/(\epsilon M_{\rm p}^2)$.\footnote{The sound speed is slightly superluminal due to the presence of the gauge field background. This kind of superluminality is common in ``k-essence" theories on classical backgrounds and does not cause the causal paradoxes \cite{Babichev:2007dw}.} In our effective theory, however, the solution~(\ref{jsol}) gives ${\cal G}_{\rm s} \simeq 4a_2X^2/H^2 \simeq 12 M_{\rm p}^2$ and $c_{\rm s} \simeq \sqrt{\epsilon_{\rm s}/12}$, where (\ref{rho}) and $X \simeq -a_1/(2a_2)$ have been used. 

The power spectrum of curvature perturbations is given by \cite{Kobayashi:2010cm, Kobayashi:2011nu}
\begin{align}\label{powerzeta}
{\cal P}_{\zeta}\equiv \frac{k^3}{2\pi^2}|\zeta_{\rm k}|^2 \simeq \frac{H^2}{8\pi^2{\cal G}_{\rm s} c_{\rm s}^3} \ .
\end{align}
If we match the scalar power spectrum with the CMB data, ${\cal P}_{\zeta} = 2\times 10^{-9}$ \cite{Hinshaw:2012aka,Komatsu:2014ioa,Akrami:2018odb}, we get a relation
\begin{align}\label{hvalue}
\frac{H}{M_{\rm p}}\simeq 2.0\times 10^{-5}\left(\frac{\epsilon_{\rm s}}{10^{-2}} \right)^{3/4} \ ,
\end{align}
with ${\cal G}_{\rm s} \simeq 12M_{\rm p}^2$ and $c_{\rm s} \simeq \sqrt{\epsilon_{\rm s}/12}$.

To estimate the characteristic scale $M$, let us define $-a_1/(2a_2) \equiv M^4$ and rescale $\phi$ to $a_1 = -1$.
Then the solution~(\ref{jsol}) is approximated to $X \simeq M^4$.
Combining $3H^2M_{\rm p}^2\simeq -K \simeq M^4/2$ and (\ref{hvalue}), we get relations
\begin{align}\label{mvalue}
\frac{M}{M_{\rm p}} \simeq 7.3\times 10^{-3} \left(\frac{\epsilon_{\rm s}}{10^{-2}} \right)^{3/8} \ ,\quad
\frac{H}{M} \simeq 2.8\times 10^{-3} \left(\frac{\epsilon_{\rm s}}{10^{-2}} \right)^{3/8} \ ,
\end{align}
which are consistent with our effective theory unless $a_3$ and $\lambda/f$ are fine-tuned to make $\epsilon_{\rm s}$ very small.

We can find a constraint on $\lambda$ in terms of $m_Q$ or $\xi$. From the definition of $\xi$ and relations~(\ref{mvalue}), we get
\begin{align}
\frac{f}{M_{\rm p}} \simeq 1.8 \frac{\lambda}{\xi} \ .
\end{align}
Using $H < f < M_{\rm p}$ and (\ref{hvalue}), we find 
\begin{align}
\label{eq:lambda}
 1.1\times 10^{-5} \left( \frac{\epsilon_{\rm s}}{10^{-2}}\right)^{3/4} < \frac{\lambda}{\xi} <  0.56 \ .
\end{align}
The stationary condition~(\ref{dphistar}) relates $\xi$ to $m_Q$ as $\xi = m_Q +1/m_Q -\epsilon/(2m_Q)$. The constraint from backreaction of particle production by the gauge field demands $m_Q=$ a few at most (section~\ref{sec:back}); thus, $\lambda$ cannot be much larger than unity. This is in stark contrast with ``Chromo-natural inflation'' \cite{Adshead:2012kp}, for which $\lambda\gg 1$ is required. Such a large coupling is not expected for an axion \cite{Agrawal:2018mkd}. On the other hand, our setup allows $\lambda$ to be more compatible with standard scenarios such as the KSVZ axion \cite{Kim:1979if,Shifman:1979if}.

The spectral tilt of (\ref{powerzeta}) is given by
\begin{align}
n_{\rm s} - 1 &\equiv \left.\frac{d\ln{ {\cal P}_{\zeta} }}{d\ln{k}} \right|_{c_{\rm s}k=aH} 
\simeq -2\epsilon - g_{\rm s} -3 \delta_{\rm s} \ , \\
g_{\rm s} &\equiv \frac{\dot{\cal G}_{\rm s}}{H{\cal G}_{\rm s}} \ ,\quad 
\delta_{\rm s} \equiv \frac{\dot{c}_{\rm s}}{Hc_{\rm s}} \ . \nonumber
\end{align}
The precise value of the tilt depends on details of shift symmetry breaking terms (potential, field-dependent coefficients, and backreaction).
Barring cancellations among terms due to fine-tuning, we expect $\epsilon \sim \epsilon_{\rm s} \sim {\cal O}(10^{-2})$ to match the tilt with CMB observations. Therefore, while the potential-driven axion-SU(2) model of ``Chromo-natural inflation'' \cite{Adshead:2012kp} and $(F\tilde{F})^2$-driven ``Gaugeflation'' model \cite{Maleknejad:2011jw, Maleknejad:2011sq} are ruled out observationally by their predicted values of $n_{\rm s}$ and the tensor-to-scalar ratio $r$ \cite{Adshead:2013qp,Adshead:2013nka,Namba:2013kia}, our construction can be made compatible with observations. 

We have a tight constraint on $\epsilon_{\rm s}$ from the scalar bispectrum. The non-linear parameters of equilateral and orthogonal scalar non-Gaussianities can be estimated as $f_{\rm NL}^{\rm equil} \sim f_{\rm NL}^{\rm ortho} \sim 0.1/c_{\rm s}^2 \sim 1/\epsilon_{\rm s}$ in the absence of the SU(2) field background \cite{Mizuno:2010ag, DeFelice:2011uc, DeFelice:2013ar}. 
Assuming that scalar non-Gaussianity of our model is dominated by the vacuum fluctuation, we can compare this prediction with the constraint from the CMB data of Planck \cite{Akrami:2019izv}, $|f_{\rm NL}^{\rm equil}| \sim |f_{\rm NL}^{\rm ortho}| \lesssim 100$. We then find a lower bound $\epsilon_{\rm s} \gtrsim 10^{-2}$.

This estimate might change when we take into account non-linearity in the gauge field perturbation \cite{Papageorgiou:2018rfx,Papageorgiou:2019ecb}. Adding non-Gaussian contribution to the scalar perturbation from the gauge field would increase the lower bound for $\epsilon_{\rm s}$ which, in turn, increases the lower bound for the tensor-to-scalar ratio presented in the next section.

\subsection{Tensor perturbation: Tensor-to-scalar ratio, chirality, and non-Gaussianity}
For tensor perturbations, the gauge field contribution affects the observable signal of the primordial gravitational wave significantly due to instability of the tensor mode of the gauge field perturbation shortly before the horizon exit \cite{Adshead:2013qp,Adshead:2013nka,Maleknejad:2012fw,Dimastrogiovanni:2012ew}.

We write gauge tensor perturbations as $\delta A_i^a = \delta_j^a a(t) T_{ij}(t,{\bf x})$, where $T_
{ij}$ is a symmetric, traceless and divergence-free tensor.\footnote{This variable $T_{ij}$ is related to those in the literature as follows: $t_{ij}=aT_{ij}$ in \cite{Maleknejad:2011jw, Maleknejad:2011sq,Adshead:2013qp,Adshead:2013nka,Obata:2016tmo,Dimastrogiovanni:2016fuu}, $t_{ij}=T_{ij}$ in \cite{Dimastrogiovanni:2012ew}, 
$\tilde{\gamma}_{ij}=T_{ij}$ in \cite{Maleknejad:2016qjz}, and $B_{ij}=T_{ij}/M_{\rm p}$ in \cite{Maleknejad:2018nxz}.} Tensor perturbations are invariant under both coordinate and SU(2) gauge transformations at linear order
\cite{Maleknejad:2011jw, Maleknejad:2011sq}. 
We find the tensor quadratic action at leading order as
\begin{align}\label{lag_h2}
S_{h^2} &= \int d^3xdt\  a^3\frac{{\cal G}_{\rm t}}{8}\left[ \dot{h}_{ij}^2 -\frac{c_{\rm t}^2}{a^2}(\partial_k h_{ij})^2\right]\ , \\
{\cal G}_{\rm t} &=  M_{\rm p}^2 -2a_4X \ ,\quad
c_{\rm t}^2 = \frac{1}{{\cal G}_{\rm t} }\left(M_{\rm p}^2+2a_4X\right) \ , \nonumber\\
S_{h T} & = \int d^3xdt\ a^3 HQ h_{ij}\left(\dot{T}_{ij}+\frac{m_Q}{a}\epsilon^{ikl}\partial_kT_{jl}\right) \ , \\
S_{T^2} &= \int d^3xdt\ a^3 \frac12 \left[ \dot{T}_{ij}^2 -\frac{1}{a^2}(\partial_kT_{ij})^2- 2\xi m_Q H^2T_{ij}^2 +2(\xi+m_Q)\frac{H}{a}\epsilon^{ijk}T_{k l}\partial_iT_{jl}\right] \ ,\nonumber
\end{align}
where $m_Q=g_AQ/H$ and $\xi=\lambda\dot\phi/(2fH)$.
The gravitational sector (\ref{lag_h2}) is modified only by the kinetic term with $a_4$ and was derived, {\it e.g.}, in \cite{Germani:2011ua}.
This contribution is sub-leading to that from general relativity in our construction, and it modifies slightly the canonical normalization and speed of gravitational waves, $c_{\rm t}$.\footnote{The propagation speed of gravitational waves can be either subluminal or superluminal depending of the sign of $a_4$. The subluminality/superluminality can be removed by rescaling the time coordinate and does not change causal structure; it does not change observable quantities for long-wave modes \cite{Creminelli:2014wna, Watanabe:2015uqa, Motohashi:2015pra, Domenech:2015hka}. For short-wave modes, there could be emission of gravitons by gravitational Cerenkov radiation for $c_{\rm t}< 1$ \cite{Kimura:2011qn} or that of photons (SU(2) particles) by Cerenkov radiation for $c_{\rm t}>1$ \cite{Brax:2015dma}.}

\begin{figure}
\centering
\includegraphics[width=0.8\textwidth]{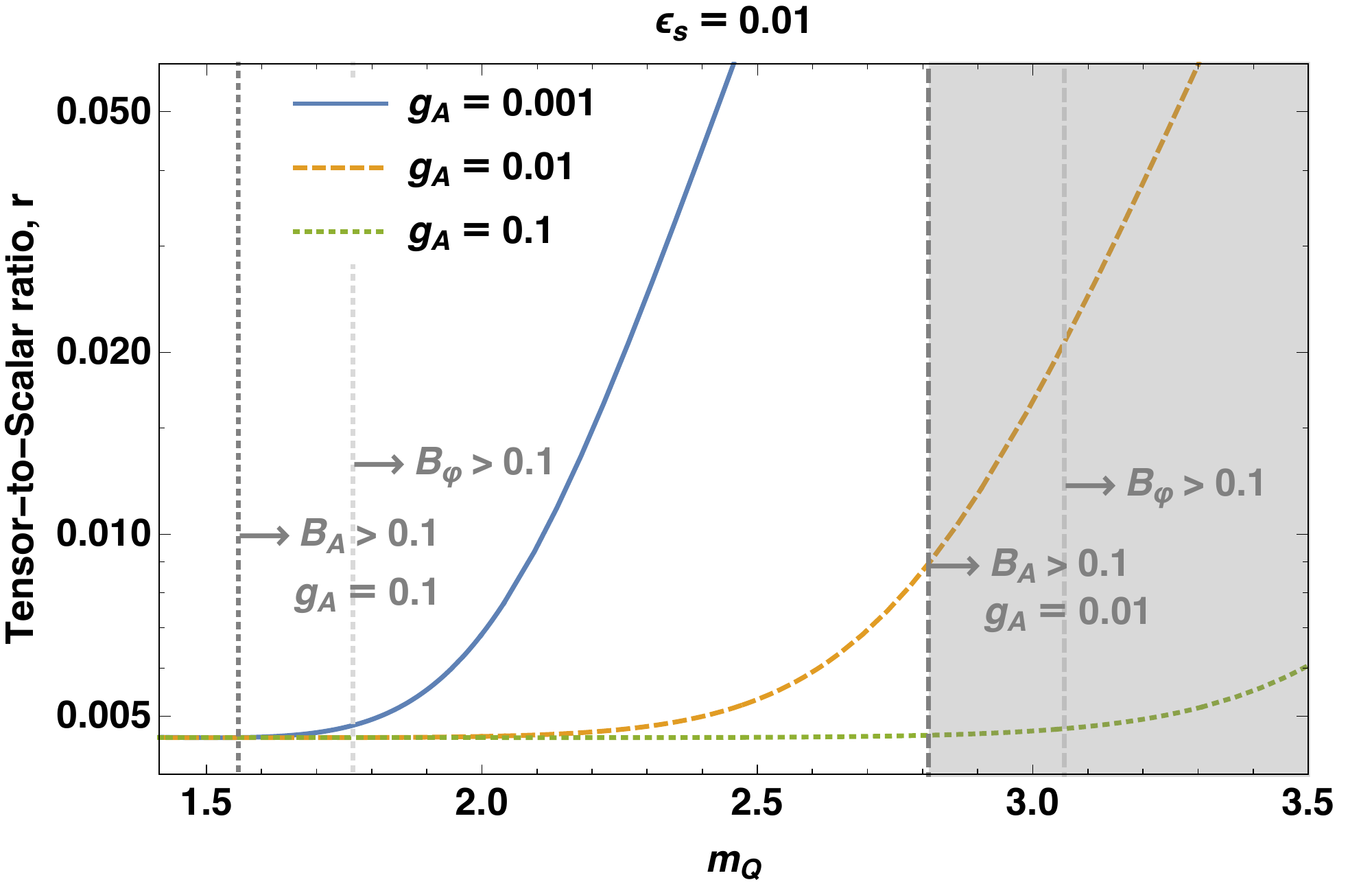}
\caption{Tensor-to-scalar ratio, $r$, as a function of the dimensionless gauge field mass, $m_Q\equiv g_AQ/H$, for different values of gauge coupling constants, $g_A$, and $\epsilon_{\rm s}=0.01$. The shaded region $m_Q \gtrsim 2.8$ indicates that the backreaction on the energy density by spin-2 particle production of the gauge field is sizable and the linear perturbation analysis cannot be trusted. For $g_A = 0.01$, the backreaction on the $Q$ field equation is also sizable in the region where $m_Q \gtrsim 2.8$. For $g_A =0.1$, the backreaction on the $Q$ field equation is sizable in the region where $m_Q \gtrsim 1.6$. See section~\ref{sec:back} for the precise meaning of the shaded region and vertical lines. \label{fig1}}
\end{figure}

Following the method of \cite{Maleknejad:2016qjz,Maleknejad:2018nxz}, we calculate the tensor-to-scalar ratio as 
\begin{align}\label{tratio}
r &= 16\frac{{\cal G}_{\rm s}c_{\rm s}^3}{{\cal G}_{\rm t}c_{\rm t}^3}\left[ 1+ \frac{Q^2e^{\pi (\xi+m_Q)}|G_+|^2}{2M_{\rm p}^2} \right] 
\simeq 4.6\times 10^{-3} \left(\frac{\epsilon_{\rm s}}{10^{-2}}\right)^{3/2} \left[ 1+ \frac{H^2m_Q^2e^{\pi (\xi+m_Q)}|G_+|^2}{2g_A^2M_{\rm p}^2} \right] \ , 
\end{align}
with ${\cal G}_{\rm s}\simeq 12 M_{\rm p}^2$, $c_{\rm s}\simeq \sqrt{\epsilon_{\rm s} /12}$, ${\cal G}_
{\rm t} \simeq M_{\rm p}^2$, and $c_{\rm t} \simeq 1$. The first term is the usual vacuum contribution \cite{Grishchuk:1974ny,Starobinsky:1979ty}, while the second term is the gauge field contribution. 
Note that $|G_+|^2=|G_+(m_Q)|^2 \lesssim {\cal O}(10^{-3})$, whose exact expression can be found in equation (E.6) of \cite{Maleknejad:2018nxz}. 
Only one of the $\pm 2$ helicity states is amplified shortly before the horizon exit, resulting in a chiral gravitational wave signal \cite{Adshead:2013qp,Adshead:2013nka,Maleknejad:2012fw,Dimastrogiovanni:2012ew}.
For $\xi>0$ (hence $\lambda\dot\phi>0$), the $+2$ helicity state grows exponentially while the $-2$ helicity state stays at the same level as the vacuum fluctuations. 

Ignoring the gauge field contribution, we find $r\gtrsim 5\times 10^{-3}$ for $\epsilon_{\rm s}\gtrsim 0.01$ given by the constraint on scalar non-Gaussianity. The gauge field contribution further increases the tensor-to-scalar ratio as shown in figure~\ref{fig1}. Here, we show $r$ as a function of $m_Q$ and $g_A$ for $\epsilon_{\rm s}=0.01$.
Since the stationary condition~(\ref{dphistar}) gives $\xi = m_Q +1/m_Q -\epsilon/(2m_Q)$ and $H$ is fixed  by (\ref{hvalue}), $r$ is solely specified by $m_Q$ if $g_A$ and $\epsilon_{\rm s}$ are provided. 
The smaller $g_A$ is, the more sensitive to $m_Q$ the amplification of the tensor mode of the gauge field becomes. This is because a small $g_A$ gives a large gauge field value $Q$ for a given $m_Q$.
For $g_A = 10^{-3}$ (blue line), $m_Q \lesssim 2.5$ is compatible with the observational constraint $r < 0.06$ \cite{Ade:2018gkx}.
For $g_A = 10^{-2}$ (orange dashed line), $m_Q \lesssim 3.3$ is compatible with $r < 0.06$.

\begin{figure}
\centering
\includegraphics[width=0.8\textwidth]{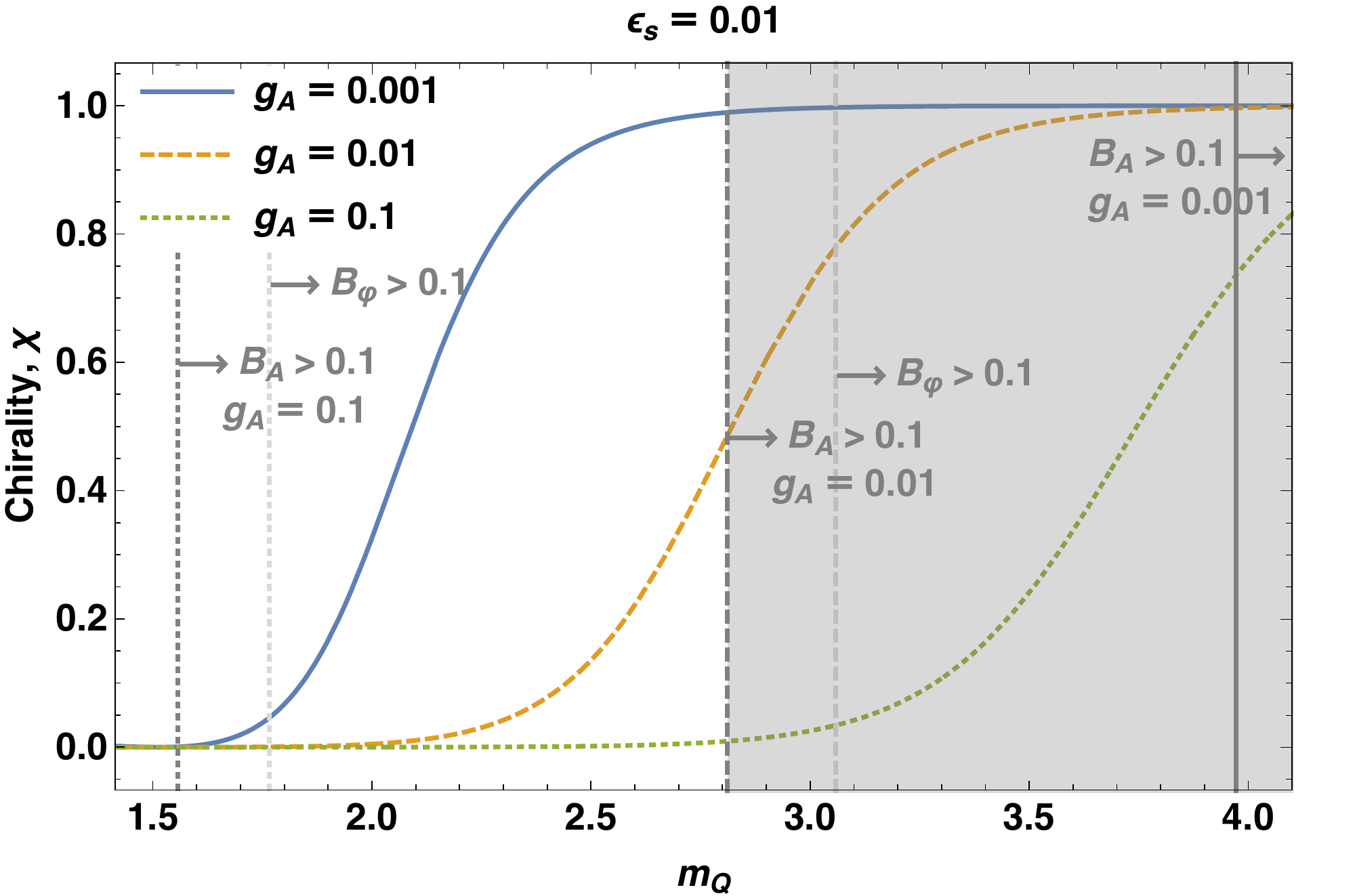}
\caption{Chirality, $\chi$, as a function of $m_Q$ for different $g_A$ and $\epsilon_{\rm s}=0.01$. The shaded region and vertical lines are the same as in figure~\ref{fig1}. \label{fig2}}
\end{figure}

We calculate chirality of the gravitational wave as 
$\chi = {r_{\rm sourced}}/({r_{\rm vacuum}+r_{\rm sourced}})$,
where $r_{\rm vacuum}$ and $r_{\rm sourced}$ are given by the first and second terms in (\ref{tratio}), respectively.
We show $\chi$ as a function of $m_Q$ and $g_A$ for $\epsilon_{\rm s}=0.01$ in figure~\ref{fig2}.

The self-coupling of the SU(2) gauge field generates the tensor bispectrum at tree level
\cite{Agrawal:2017awz,Agrawal:2018mrg}. The  bispectrum of the $+2$ helicity state of the primordial gravitational wave at the equilateral configuration is given by $B_{h,\rm sourced}^{RRR}(k,k,k)/[P_{h,\rm sourced}^R(k)]^2\simeq 1.816\exp(0.841m_Q)/\epsilon_B$ \cite{Agrawal:2018mrg}, where $\epsilon_B\equiv g_A^2Q^4/(H^2M_{\rm p}^2)=m_Q^2(Q/M_{\rm p})^2\ll 1$ and ``$R$'' stands for the right-handed ($+2$) helicity state. This formula is accurate for $3\lesssim m_Q\lesssim 5$. 
This is much larger than that of the vacuum contribution at the same configuration, $B_{h,\rm vacuum}^{RRR}(k,k,k)/[P_{h,\rm vacuum}^R(k)]^2\simeq 3.586$ \cite{Maldacena:2011nz,Agrawal:2018gzp}. The total bispectrum of the $+2$ helicity state is therefore given by
\begin{eqnarray}
\nonumber
\frac{B_{h}^{RRR}(k,k,k)}{[P_{h}^R(k)]^2}&=&\frac{B_{h,\rm vacuum}^{RRR}(k,k,k)+B_{h,\rm sourced}^{RRR}(k,k,k)}{[P_{h,\rm vacuum}^R(k)+P_{h,\rm sourced}^R(k)]^2}\\
&\approx& 3.586(1-f_{\rm s})^2+1.816f_{\rm s}^2\exp(0.841m_Q)/\epsilon_B\,,
\label{eq:appbispec}
\end{eqnarray} 
where $f_{\rm s}\equiv P_{h,\rm sourced}^R/(P_{h,\rm vacuum}^R+P_{h,\rm sourced}^R)=r_{\rm sourced}/(r_{\rm vacuum}/2+r_{\rm sourced})$ is the fraction of the sourced power spectrum in the total right-handed gravitational wave power spectrum. Thus, the bispectrum can be a powerful probe of the  gravitational wave sourced by the SU(2) gauge field \cite{Agrawal:2017awz,Agrawal:2018mrg,Shiraishi:2019yux}. 

\begin{figure}
\centering
\includegraphics[width=0.8\textwidth]{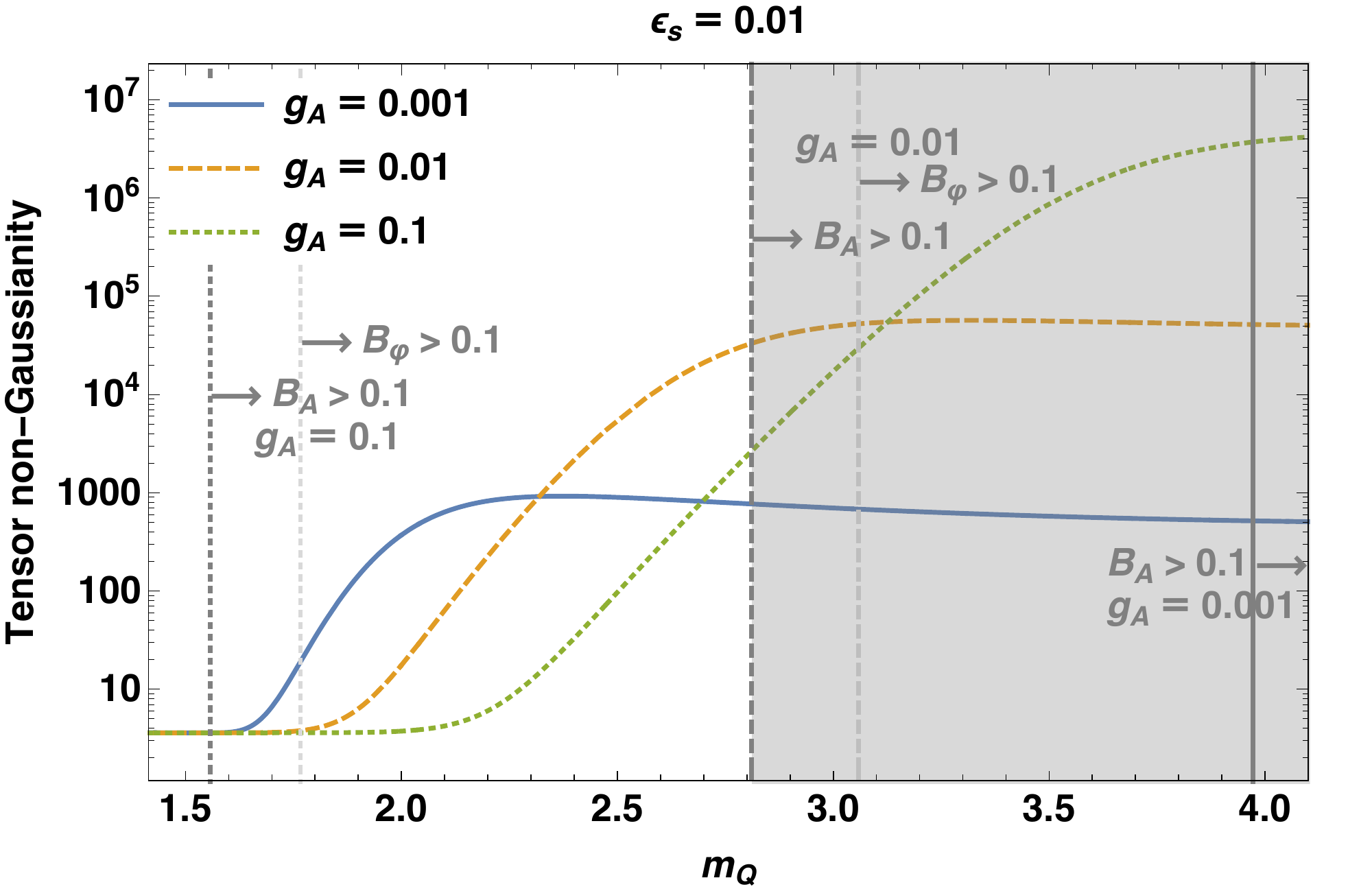}
\caption{Tensor non-Gaussianity of the $+2$ helicity state at the equilateral configuration, $B_{h}^{RRR}(k,k,k)/[P_{h}^R(k)]^2$ (\ref{eq:appbispec}), as a function of $m_Q$ for different $g_A$ and $\epsilon_{\rm s}=0.01$. The shaded region and vertical lines are the same as in figure~\ref{fig1}. \label{fig3}}
\end{figure}

We show tensor non-Gaussianity at the equilateral configuration, $B_{h}^{RRR}(k,k,k)/[P_{h}^R(k)]^2$, as a function of $m_Q$ in figure~\ref{fig3}. While (\ref{eq:appbispec}) is accurate for $3\lesssim m_Q\lesssim 5$, we use it for lower $m_Q$ as well. Figure 9 of ref.~\cite{Agrawal:2018mrg} suggests that the formula overestimates
tensor non-Gaussianity by a factor of two at $m_Q=2$; thus, the lines in $m_Q<2$ before reaching the vacuum level (plateau) should be regarded as an order of magnitude estimate, as the actual values can be smaller than the lines by a factor of several.

\subsection{Backreaction of particle production}\label{sec:back}
Due to significant particle production of gauge tensor perturbations with $+2$ helicity state $T_{+}(k)$, this mode backreacts on the background energy density and field equations for axion and gauge fields \cite{Dimastrogiovanni:2016fuu,Fujita:2017jwq,Maleknejad:2018nxz}. In our setup, this effect is much stronger than the Schwinger process in which quantum fields are sourced by the background gauge field \cite{Lozanov:2018kpk, Domcke:2018gfr, Mirzagholi:2019jeb,Maleknejad:2019hdr}.
Validity of the linear perturbation theory analysis given in this paper requires the backreaction terms be much smaller than the total energy density perturbation and the other terms in the field equations.

The regularized energy density fraction in the gauge tensor perturbations $T_{ij}$ is given by
\begin{align}
\frac{\langle\delta\rho_T \rangle_{\rm reg}}{\rho} = \frac{(6m_Q +2/m_Q)H^2}{12 \pi^2 M_{\rm p}^2}{\cal K_{\rm reg}}\left[\frac{3m_Q^2+5}{6m_Q+2/m_Q} \right] \ ,
\end{align}
where regularization is done by adiabatic subtraction and the expression of ${\cal K}_{\rm reg}[x]$ is given in equation~(4.11) of \cite{Maleknejad:2018nxz}.
The shaded region of $m_Q\gtrsim 2.8$ in figures~\ref{fig1}, \ref{fig2} and \ref{fig3} shows $|\langle\delta\rho_T \rangle_{\rm reg}|/\rho > 10^{-6}$ where the linear analysis cannot be trusted.
The constraint is weaker for smaller values of $H$. 
The relation~(\ref{hvalue}) means that smaller values of $H$ correspond to smaller $\epsilon_{\rm s}$, which has a lower bound from scalar non-Gaussianity. Therefore, backreaction on the energy density shown in the figures is a lower bound.  
Note that $\langle \delta\rho_T\rangle_{\rm reg}$ is negative-definite since the particle production occurs due to the tachyonic mass of $T_+(k)$.

Other measures of the backreaction are given by \cite{Maleknejad:2018nxz}
\begin{align}\label{backreaction}
\langle {\cal J}_A\rangle_{\rm reg} &= \frac{g_AH^3}{6\pi^2}{\cal K}_{\rm reg}\left[m_Q+\frac{1}{m_Q}\right] \ ,\quad
\langle {\cal P}_{\phi}\rangle_{\rm reg} = \frac{3\lambda H^4}{4\pi^2f}{\cal K}_{\rm reg}[m_Q] \ , \\
B_A &\equiv \frac{\langle {\cal J}_A \rangle_{\rm reg}}{H^2Q} = \frac{g_A^2}{6\pi^2 m_Q}{\cal K}_{\rm reg}\left[m_Q+\frac{1}{m_Q}\right] \ , \\
B_{\phi} &\equiv \frac{\langle {\cal P}_{\phi}\rangle_{\rm reg}}{\lambda g_A HQ^3/f}= \frac{3g_A^2}{4\pi^2m_Q^3}{\cal K}_{\rm reg}[m_Q] \ ,
\end{align}
where the backraction terms~(\ref{backreaction}) appear as corrections to the field equations of motion for $Q$ (\ref{eq_Q}) and $\phi$ (\ref{eq_phi}), respectively. 

The effective potential for $Q$ (\ref{eq:effpot}) may be modified due to backreaction as
\begin{align}
U_{\rm eff}(Q) = \frac12(\dot{H}+2H^2)Q^2 + \frac12 g_A^2Q^4 - \frac{g_A\lambda}{3f}\dot{\phi}Q^3 -\langle {\cal J}_A\rangle_{\rm reg}Q, 
\end{align}
where $\langle {\cal J}_A\rangle_{\rm reg}$ in the last term is positive-definite and does not spoil the existence of a nontrivial value of $Q$ for the stationary state even if the magnitude of $\langle {\cal J}_A\rangle_{\rm reg}Q$ is sizable. However, the linear perturbation analysis is not reliable when $B_A > 0.1$, which excludes most of the parameter space for $g_A \gtrsim 0.1$. We thus need more careful analysis in this parameter space. The gray vertical lines in figures~\ref{fig1}, \ref{fig2} and \ref{fig3} show $B_A=0.1$ for $g_A=0.1$ (dotted, at $m_Q\simeq 1.56$), $g_A=0.01$ (dashed, at $m_Q\simeq 2.81$) and $g_A=0.001$ (solid, at $m_Q \simeq 3.97$). Backreaction on the background gauge field equation of motion cannot be ignored in the parameter space right to these lines.

Since $\langle {\cal P}_{\phi}\rangle_{\rm reg}$ is also positive-definite, it contributes to the shift symmetry breaking term $K_{,\phi}$ in (\ref{eq_phi}). In this case, the first iterative solution (\ref{jsol}) is modified to
\begin{align}\label{jsolmod}
J^{(1)} = -\frac{g_A\lambda}{f}Q^3 +\frac{K_{,\phi} +\langle{\cal P}_{\phi}\rangle_{\rm reg}}{3H}+ \frac{C}{a^3} \ .
\end{align}
As a result, (\ref{epsilonstar}) is changed to $\epsilon_* \approx \dot\phi(K_{,\phi}+\langle{\cal P}_{\phi}\rangle_{\rm reg})/(6H^3M_{\rm p}^2)$. 
In the absence of $K_{,\phi}$, the effect becomes $\epsilon_* \simeq B_{\phi}\xi m_Q Q^2/(3 M_{\rm p}^2) = \xi {\cal K}_{\rm reg}[m_Q]H^2/(4\pi^2 M_{\rm p}^2)$. For instance, $\epsilon_* = 9.4\times 10^3 H^2/M_{\rm p}^2$ for $m_Q=2.8$ and smaller for smaller $m_Q$; it is too small to account for the tilt of the curvature power spectrum.
The magnitude of the right hand side of (\ref{jsolmod}) does not affect the existence of a nontrivial solution of $\dot\phi$.
However, the linear perturbation analysis is not reliable when $B_{\phi}>0.1$.
The light gray vertical lines in figures~\ref{fig1}, \ref{fig2} and \ref{fig3} show $B_\phi=0.1$ for $g_A=0.1$ (dotted, at $m_Q\simeq 1.77$), $g_A=0.01$ (dashed, at $m_Q\simeq 3.06$) and $g_A=0.001$ (not shown, at $m_Q\simeq 4.36$).
Backreaction on the background axion field equation of motion cannot be ignored in the parameter space right to these lines, which is always weaker than $B_A>0.1$ by roughly an order of magnitude.

\section{Conclusion}\label{sec:conclusion}
In this paper, we have constructed a low-energy effective Lagrangian (\ref{action}) for a pseudo scalar (axion) field with shift symmetry, which contains no more than four spacetime derivatives. We have coupled the axion field to SU(2) gauge fields via a Chern-Simons coupling. Focusing on a class of inflationary models driven by kinetic terms (rather than by a potential), we have obtained the solutions to the background equations of motion with softly broken shift symmetry.

The scalar curvature perturbation is non-Gaussian when the speed of sound parameter $c_{\rm s}\simeq \sqrt{\epsilon_{\rm s}/12}$ is small. Using the observational constraint on scalar non-Gaussianity from the CMB data, we find a lower bound for $\epsilon_{\rm s}$ which, in turn, yields a lower bound for the tensor-to-scalar ratio of the primordial gravitational wave, $r\gtrsim 5\times 10^{-3}$, from the vacuum fluctuation (the first term in (\ref{tratio})). This is within the reach of upcoming ground-based \cite{Abazajian:2016yjj,Ade:2018sbj} and space-borne CMB experiments \cite{Sugai:2020pjw,Hanany:2019wrm}. 

The contribution from the tensor perturbation in the SU(2) gauge field further increases $r$ (the second term in (\ref{tratio}); figure~\ref{fig1}). As this contribution is chiral \cite{Adshead:2013qp,Adshead:2013nka,Maleknejad:2012fw,Dimastrogiovanni:2012ew} and non-Gaussian \cite{Agrawal:2017awz,Agrawal:2018mrg,Fujita:2018vmv,Dimastrogiovanni:2018xnn}, it makes the total primordial gravitational wave partially chiral (figure~\ref{fig2}) and non-Gaussian (figure~\ref{fig3}). This tensor non-Gaussianity can be probed by CMB experiments (see \cite{Shiraishi:2019yux} for a recent review). Chirality can also be probed by CMB experiments as well as by laser interferometers in a suitable configuration (e.g., \cite{Thorne:2017jft}). These predictions are distinct from nearly Gaussian and non-chiral gravitational waves of the vacuum fluctuation; thus, prospects for distinguishing between the sourced and vacuum contributions are good. An added bonus is that chiral gravitational waves can generate the baryon number in the Universe via gravitational anomaly in the lepton number current \cite{Alexander:2004us,Lyth:2005jf,Fischler:2007tj}.
Estimates of the baryon number from non-Abelian gauge fields are given in \cite{Maleknejad:2014wsa,Maleknejad:2016dci,Noorbala:2012fh,Caldwell:2017chz}.

The tilt of the curvature power spectrum, $n_{\rm s}$, is generated via softly broken shift symmetry. 
This can be achieved by a subdominant potential and $\phi$-dependent coefficients $a_i(\phi)$.
Therefore, our model can be made compatible with the CMB data, whereas the original inflation models based on the SU(2) gauge field, ``Gaugeflation'' \cite{Maleknejad:2011jw, Maleknejad:2011sq} and ``Chromo-natural inflation'' \cite{Adshead:2012kp}, have been ruled out by the constraints on $n_{\rm s}$ and $r$.

Another issue of the Chromo-natural inflation is that it requires $\lambda\gg 1$ for successful phenomenology, which is in tension with standard constructions of axion models \cite{Agrawal:2018mkd}. We find that our model allows $\lambda\ll 1$ (\ref{eq:lambda}).

We thus conclude that our effective Lagrangian (\ref{action}) can yield well-motivated inflationary models which are phenomenologically viable and predict distinct properties of scalar and tensor perturbations; namely, tilted and non-Gaussian scalar perturbations and partially chiral and non-Gaussian primordial gravitational waves.

Finally, we comment on reheating scenarios in our construction.
They are dependent on how shift symmetry is broken. If the scalar spectral tilt is produced by a potential, inflation may end when the potential becomes too steep to keep inflation.
In this case, reheating processes would proceed via particle production in potential energy domination.
If the scalar spectral tilt is produced by $\phi$-dependence in the kinetic terms, inflation may end when the (nearly) constant speed solution disappears. 
In this case, reheating processes would proceed via particle production in kinetic energy domination, called ``kination" \cite{Spokoiny:1993kt}.
In any case the processes depend on the details of the shift symmetry breaking sector, and we leave this interesting question for future work.

\section*{Acknowledgments}
We thank Kaloian Lozanov and Azadeh Maleknejad for collaboration in the early phase and valuable discussions. We also thank Emanuela Dimastrogiovanni, Valerie Domcke, Matteo Fasiello for clarifying their work and Giovanni Cabass, Elisa Ferreira, Fabio Finelli, Raphael Flauger, Kohei Kamada, Leila Mirzagholi, Ryo Namba, and Filippo Vernizzi for useful conversations.
This research was supported in part by the Excellence Cluster ORIGINS which is funded by the Deutsche Forschungsgemeinschaft (DFG, German Research Foundation) under Germany's Excellence Strategy -- EXC-2094 -- 390783311.
YW acknowledges support from JSPS KAKENHI Grant No.~JP16K17712.


\end{document}